\global\def\draftcontrol{0}

   \def\versionno{Vogel CS}

\catcode`\@=11

\expandafter\ifx\csname draftcontrol\endcsname\relax\global\def\draftcontrol{0} 
\fi 

{\count255=\time\divide\count255 by 60 
\xdef\hourmin{\number\count255} 
\multiply\count255 by-60\advance\count255 by\time 
\xdef\hourmin{\hourmin:\ifnum\count255<10 0\fi\the\count255}} 
\def\draftdate{\number\month/\number\day/\number\year\ \ \ \hourmin } 

\newcommand\makepapertitle{\par

  \begingroup 
    \renewcommand\thefootnote{\@fnsymbol\c@footnote}%
    \def\@makefnmark{\rlap{\@textsuperscript{\normalfont\@thefnmark}}}%
    \long\def\@makefntext##1{\parindent 1em\noindent 
            \hb@xt@1.8em{%
                \hss\@textsuperscript{\normalfont\@thefnmark}}##1}%
     \newpage 
     \global\@topnum\z@   
     \@makepapertitle 
     \thispagestyle{empty}\@thanks 
  \endgroup 
  \setcounter{footnote}{0}%
  \global\let\thanks\relax 
  \global\let\makepapertitle\relax 
  \global\let\@makepapertitle\relax 
  \global\let\@thanks\@empty 
  \global\let\@author\@empty 
  \global\let\@date\@empty 
  \global\let\@title\@empty 
  \global\let\title\relax 
  \global\let\author\relax 
  \global\let\date\relax 
  \global\let\and\relax 
  \def\version{\let\version\@version\@gobble} 
} 
\def\@makepapertitle{%
  \newpage 
   \ifnum\draftcontrol=1 {} 
   \version\versionno 
   \vskip 5.5em%
   \else 
   \hfill\hbox to 3cm {\parbox{4.5cm}{\@pubnum}\hss}%
   \vskip 6.5em%
   \fi 
   \begin{center}%
   \let \footnote \thanks 
      {\hskip -0\textwidth \hbox to 1\textwidth%
        {\centerline{\Large\bf{\noindent\@title}}}}%
     \vskip 2em%
     {\normalsize
       \lineskip .5em%
       \begin{tabular}[t]{c}%
         \@author 
       \end{tabular}\par}%
     \vskip 1.5em%
     {\@bstract}%
     \end{center}%
     \vfill
     \@date%
     \vskip 1.5em%
   \par 
} 

\gdef\@pubnum{} 
\def\pubnum#1{%
  \gdef\@pubnum{#1}} 

\gdef\@bstract{} 
\def\Abstract#1{%
  \gdef\@bstract{%
   \parbox{\textwidth-0pc}{%
   \centerline{\bf Abstract}\penalty1000 
   \noindent
   \renewcommand\baselinestretch{1.0} 
   {#1}}} 
} 

\gdef\@email{}
\def\email#1{%
   \gdef\@email{%
   Email: {\tt #1}}
}

\def\ps@paper{\let\@mkboth\@gobbletwo%
     \ifnum\draftcontrol=1 
        \def\@oddfoot{\hbox to \textwidth{\tiny \versionno \hfil\tiny\draftdate}%
        \hskip -\textwidth \hbox to \textwidth{\hfil\rm\thepage\hfil}}%
     \else\def\@oddfoot{\hbox to \textwidth{\hfil\rm\thepage\hfil}} 
     \fi 
     \let\@evenfoot\@oddfoot 
} 

\def\body{\clearpage 
          \pagestyle{paper} 
        } 
\newenvironment{acknowledgments}{%
\vskip 3.25ex 
\addcontentsline{toc}{section}{Acknowledgments}
\noindent {\bf Acknowledgments} 
} 

\def\@version#1{\ifnum\draftcontrol=1 
\typeout{}\typeout{#1}\typeout{} 
\vskip3mm\centerline{\hbox{\fbox{\normalsize{\tt DRAFT -- #1 -- } 
                   {\draftdate}}}}\vskip3mm 
\fi} 
\let\version\@version 
\long\def\eqlabel#1{\ifnum\draftcontrol=1 
                    \tag@false  
                    \tag*{(\theequation) \hbox to -0.2cm{\hspace{0cm}\small{#1}\hss}} 
                    \refstepcounter{equation}  
                    \edef\@currentlabel{\theequation} 
                    \ltx@label{#1}          
                    \else 
                    \label{#1} 
                    \fi 
                    } 
\let\st@bibitem\@bibitem 
\let\st@lbibitem\@lbibitem 
\ifnum\draftcontrol=1 
  \def\@bibitem#1{%
    \st@bibitem{#1}\a@@label{#1}\ignorespaces} 
  \def\@lbibitem[#1]#2{%
    \st@lbibitem[#1]{#2}\a@@label{#2}\ignorespaces} 
  \def\a@@label#1{%
    \gdef\a@lab{\smash{\normalfont\small#1}} 
    \ifvmode 
      \if@inlabel 
        \global\setbox\@labels\hbox{%
          \llap{\a@lab\let\a@lab\relax 
                \kern\@totalleftmargin\kern\marginparsep}%
          \box\@labels}%
      \fi 
    \fi} 
\fi 

\documentclass[12pt,letterpaper]{article} 

\usepackage{amsmath,bm,amsfonts,amssymb,array,calc,amsthm,rotating}
\usepackage{epsfig,psfrag} 
\usepackage{rotating}
\usepackage{amscd}
\usepackage{graphicx}
\usepackage{color}
\usepackage[colorlinks=false]{hyperref}

\renewcommand\baselinestretch{1.25} 
\renewcommand\section{\@startsection {section}{1}{\z@}%
                                   {-3.5ex \@plus -1ex \@minus -.2ex}%
                                   {2.3ex \@plus.2ex}%
                                   {\normalfont\large\bfseries}} 
\renewcommand\subsection{\@startsection{subsection}{2}{\z@}%
                                   {-3.25ex\@plus -1ex \@minus -.2ex}%
                                   {1.5ex \@plus .2ex}%
                                   {\normalfont\normalsize\bfseries}} 
\renewcommand\subsubsection{\@startsection{subsubsection}{3}{\z@}%
                                   {-3.25ex\@plus -1ex \@minus -.2ex}%
                                   {1.5ex \@plus .2ex}%
                                   {\normalfont\normalsize\it}} 
\renewcommand\paragraph{\@startsection{paragraph}{4}{\z@}%
                                   {-1.75ex\@plus -1ex \@minus -.2ex}%
                                   {1ex \@plus .2ex}%
                                   {\normalfont\normalsize\bf}} 
\renewcommand\subparagraph{\@startsection{subparagraph}{5}{\z@}%
                                   {-1.25ex\@plus -0ex \@minus -.2ex}%
                                   {-2ex \@plus .2ex}%
                                   {\normalfont\normalsize\it}}

\numberwithin{equation}{section}

\long\def\@makecaption#1#2{%
  \vskip\abovecaptionskip
  \sbox\@tempboxa{{\bf #1:} #2}%
  \ifdim \wd\@tempboxa >\hsize
    {\small\bf #1:} {\small #2}\par
  \else
    \global \@minipagefalse
    \hb@xt@\hsize{\hfil\box\@tempboxa\hfil}%
  \fi
  \vskip\belowcaptionskip}

\renewcommand*\l@section[2]{%
  \ifnum \c@tocdepth >\z@
    \addpenalty\@secpenalty
    \addvspace{.5em \@plus\p@}%
    \setlength\@tempdima{1.5em}%
    \begingroup
      \parindent \z@ \rightskip \@pnumwidth
      \parfillskip -\@pnumwidth
      \leavevmode \bfseries
      \advance\leftskip\@tempdima
      \hskip -\leftskip
      #1\nobreak\hfil \nobreak\hb@xt@\@pnumwidth{\hss #2}\par
    \endgroup
  \fi}
\renewcommand*\l@subsection{\addvspace{.0em \@plus\p@}\@dottedtocline{2}{1.5em}{2.3em}}
\renewcommand*\l@subsubsection{\addvspace{-.2em \@plus\p@}\@dottedtocline{3}{3.8em}{3.2em}}

\def\hepth#1{\href{http://xxx.arxiv.org/abs/hep-th/#1}{{arXiv:hep-th/#1}}}

\def\arxiv#1#2{\href{http://xxx.arxiv.org/abs/#1}{{arXiv:#1 [#2]}}}

\definecolor{refcol}{rgb}{0.2,0.2,0.8}
\definecolor{eqcol}{rgb}{.6,0,0}
\definecolor{purple}{cmyk}{0,1,0,0}

\gdef\@citecolor{refcol}
\gdef\@linkcolor{eqcol}
\def\colorlinkspurple{\gdef\@urlcolor{purple}}
\def\colorlinksblue{\gdef\@urlcolor{blue}}
\def\colorlinksred{\gdef\@urlcolor{red}}

\def\ie{{\it i.e.}}

\def\cf{{\it cf.}}

\def\revise#1       {\raisebox{-0em}{\rule{3pt}{1em}}%
                     \marginpar{\raisebox{.5em}{\vrule width3pt\ 
                     \vrule width0pt height 0pt depth0.5em 
                     \hbox to 0cm{\hspace{0cm}{%
                     \parbox[t]{4em}{\raggedright\footnotesize{#1}}}\hss}}}}

\def\ii           {{\it i}}

\def\Re           {{\rm Re\hskip0.1em}}

\def\sqr#1#2{{\vcenter{\vbox{\hrule height.#2pt   
 \hbox{\vrule width.#2pt height#1pt \kern#1pt 
 \vrule width.#2pt}\hrule height.#2pt}}}}

\renewcommand{\P}{\mathbb P}

\renewcommand{\S}{\mathbb S}

\newcommand{\Ical}{\mathcal I}

\newcommand{\Fcal}{\mathcal F}

\newcommand{\Ocal}{\mathcal O}

\newcommand{\Gcal}{\mathcal G}

\newcommand{\ep}{\epsilon}

\newcommand{\beq}{\begin{equation}}
\newcommand{\eq}{\end{equation}}
\newcommand{\req}[1]{(\ref{#1})}

\catcode`\@=12 

\begin{document} 


\title{Refined Chern-Simons versus Vogel universality}

\pubnum{
UCB-PTH-13/04
}
\date{April 2013}

\author{
Daniel Krefl$^a$ and Albert Schwarz$^b$  \\[0.2cm]
\it  $^a$Center for Theoretical Physics, University of California, Berkeley, USA\\
\it $^b$Department of Mathematics, University of California, Davis, USA\
}

\Abstract{
We study the relation between the partition function of refined $SU(N)$ and $SO(2N)$ Chern-Simons on the 3-sphere and  the universal Chern-Simons partition function in the sense of Mkrtchyan and Veselov. We find  a four-parameter generalization of the integral representation of universal Chern-Simons that includes  refined $SU(N)$ and $SO(2N)$ Chern-Simons for special values of parameters. The large $N$ expansion of the integral representation of refined $SU(N)$ Chern-Simons explicitly shows the replacement of the virtual Euler characteristic of the moduli space of complex curves with a refined Euler characteristic related to the radius deformed c=1 string free energy.
}

\makepapertitle

\body

\version\versionno

\vskip 1em



\section{Introduction}
In recent years, after the work of Nekrasov \cite{N02}, it became clear that there should exist a sort of refined topological string theory as a one-parameter deformation of the ordinary topological string on a local Calabi-Yau 3-fold \cite{HIV03,IKV07}. A one-parameter deformation in the sense that we treat the extra parameter not as an additional coupling constant in which we expand, but rather as a finite parameter of the theory or background, commonly denoted as $\beta$. 

This point of view differs from the original A-model definition of the refined topological string partition function as an M-theory index of spinning M2-branes, but is natural in the dual B-model \cite{KW10a,KW10b}. In detail, as the genus expansion of the B-model topological string partition function on the mirror local Calabi-Yau is entirely determined at tree-level by special geometry, and at higher loops recursively by the holomorphic anomaly equations equipped with specific boundary conditions (holomorphic ambiguity) \cite{BCOV93a,BCOV93b}, so appears to be the refined partition function, up to a $\beta$-dependent deformation of the boundary conditions.

In particular, near a conifold point in moduli space, the leading singular terms of the partition function are not anymore given by the self-dual $c=1$ string free energy, being a generating function for the virtual Euler characteristic of complex curves, as in the unrefined case, but by the radius deformed $c=1$ string free energy at radius $R=\beta$.  Correspondingly, due to the well-known CFT relation between radius deformation and orbifolding, one may see for integer $\beta>0$ refinement near the conifold point as well purely geometrically as replacing the conifold singularity by an $A_\beta$ singularity.

It has been known already for some time that ordinary $SU(N)$ Chern-Simons on $\mathbb S^3$ is dual at large $N$ to the topological string on the resolved conifold geometry, \ie, on $\Ocal(-1)\oplus\Ocal(-1)\rightarrow \P^1$ \cite{GV98}. Similarly, $SO/Sp$ Chern-Simons relates to orientifolds of the topological string on the resolved conifold \cite{SV00}. Recently, a proposal for a refined Chern-Simons partition function has been made in \cite{AS11,AS12}. In particular, for $SU(N)$ the proposed partition function yields at large $N$ the known partition function of the refined topological string on the resolved conifold, while the proposed refined $SO(2N)$ partition function has been claimed to correspond to a refined orientifold at large $N$ \cite{ASca12}.

Rather unrelated to these developments, very recently there has been a proposal for a three parameter generalization of Chern-Simons theory, dubbed universal Chern-Simons \cite{MV12,M13}. The original motivation for such a universal Chern-Simons theory goes back to the work of Vogel \cite{V11,V99}. Inspired by the analysis of Vassiliev invariants of knots Vogel introduced the notion of universal Lie algebra. This notion leads to a  prediction  that there exist knot invariants labeled by three parameters  $(a,b,c)$ that are defined up to permutation; for some specific values of these parameters we should obtain the invariants coming from Chern-Simons theory  based on simple Lie algebras.  Namely, the parameters  $(a,b,c)$ for simple Lie algebras  can be expressed in terms of eigenvalues  of the Casimir operator on the symmetric square of the Lie algebra. The eigenvalues $(a,b,c)$ can be considered as  projective coordinates on the so-called Vogel plane, the simple Lie algebras sit at special points in the Vogel plane, for classical Lie algebras they  are listed  in the table \ref{VogelTable}. 
\begin{table}
\label{VogelTable}
\begin{center}
\begin{tabular}{c|c|c|c}
$\Gcal$&$a$&$b$&$c$\\
\hline
$A_N$&-2&2&$N+1$\\
$B_N$&-2&4&$2N-3$\\
$C_N$&-2&1&$N+2$\\
$D_N$&-2&4&$2N-4$\\
\end{tabular}
\caption{Vogel parameters for the classical Lie algebras.}
\end{center}
\end{table}
The simple Lie superalgebras also sit at points in the Vogel plane. 

Notice that  Chern-Simons theory depends not only on the choice of simple Lie algebra, but also on the choice of  invariant inner product (=the choice of Casimir operator) that enters as a coupling constant $\kappa$ in the action functional. Hence the Chern-Simons theory is described by triples $(a,b,c)$ defined up to permutation, but without identification $(a,b,c)\sim (\lambda a, \lambda b, \lambda c)$. Alternatively, the Chern-Simons theory can be described by projective coordinates $(a,b,c,\kappa)$.

Remarkably, many characteristics of the Lie algebras can be expressed in universal form via Vogel's three parameters.  If we expect that there exist knot invariants depending on Vogel parameters, we should expect that  the partition functions of Chern-Simons theory for all  simple Lie algebras can be expressed in an universal fashion in terms of these parameters. This has been proved for Chern-Simons theory on $\S^3$ in \cite{MV12, M13}, where a  closed integral expression in terms of the Vogel parameters for the partition function has been derived.

Though there is no obvious relationship between refined topological string, refined Chern-Simons and the universal Chern-Simons inspired by Vogel's universal Lie algebra, one may ask if maybe the former is included at least to some extent in the latter. To take the main result ahead, in general, the refined Chern-Simons partition function on $\S^3$ is not a special case of the universal Chern-Simons partition function. However, it is natural to extend the universal Chern-Simons partition function to a function of four parameters $(a,b,c,t)$, or five parameters $(a,b,c,t,\kappa)$ in projective coordinates (``refined universal Chern-Simons"), which contains refined partition functions. 

This extension is given by the formulas of \cite{M13} where $t$ is considered as an independent parameter (\ie, we drop Vogel's condition $t=h=a+b+c$, with $h$ the dual Coxeter number). We will find that the refined Chern-Simons partition functions on $\S^3$ with refinement parameter $\beta$ of \cite{AS11,AS12} can be obtained by choosing the parameters in the $SU(N)$ case as $(-2,2\beta,\beta N,\beta N,\kappa)$ and in the $SO(2N)$ case as $(-2,4\beta,2\beta N-4\beta,2\beta N-2\beta,\kappa)$. This can be expressed universally as 
\beq
\eqlabel{bisBeta}
\boxed{
(a,b,c,\kappa)\longrightarrow (a,\beta b,\beta c,t=\beta h,\kappa)
}\,.
\eq

The outline is as follows. In section \ref{UCS} we will formulate the main claim, that is, the choice of Vogel parameters yielding integral representations of the  refined $SU(N)$ and $SO(2N)$ partition functions. Via explicit calculation of the corresponding refined partition functions, as defined in \cite{AS11,AS12}, we will proof the claim in section \ref{RCS}. The integral expression for the refined partition functions will be further used to perform a large $N$ expansion in the $SU(N)$ case in section \ref{largeN}, as this neatly illustrates that refinement essentially corresponds to replacement of the virtual Euler characteristic by a sort of refined (or parameterized, in the spirit of \cite{GHJ01}) Euler characteristic. Some technical details are collected in appendices \ref{IntApp} and \ref{TrigIdapp}.

\section{Universal Chern-Simons}
\label{UCS}
The partition function of universal Chern-Simons on $\S^3$ has been shown in \cite{MV12,M13} to be factorizable as
$$
Z^U_\Gcal=Z^U_{\rm I,\Gcal}\, Z^U_{\rm  II,\Gcal}\,.
$$
The free energy of part I reads 
$$
\Fcal_{\rm I,\Gcal}:=\log Z^U_{\rm I,\Gcal}=\int_0^\infty \frac{dx}{x(e^x-1)}\,F_\Gcal(x/\delta)\,,
$$
with
\beq
\eqlabel{FGfull}
F_\Gcal(x)=\frac{\sinh\left(\frac{x(a-2t)}{4}\right)\sinh\left(\frac{x(b-2t)}{4}\right)\sinh\left(\frac{x(c-2t)}{4}\right)}{\sinh\left(\frac{xa}{4}\right)\sinh\left(\frac{xb}{4}\right)\sinh\left(\frac{xc}{4}\right)}-\frac{(a-2t)(b-2t)(c-2t)}{abc}\,.
\eq
We further defined 
\beq
\eqlabel{DeltaDef}
\delta:=\kappa+t\,,
\eq
with $\kappa$ the Chern-Simons coupling constant and
\beq
\eqlabel{VogelRelation}
t=h=a+b+c\,,
\eq 
stands for a half of the Casimir eigenvalue of the adjoint representation, given by the sum of the Vogel parameters $a,b,c$.

Similarly, the free energy of the second part of the partition function reads
$$
\log Z^U_{\rm II,\Gcal}=\int_0^\infty \frac{dx}{x(e^x-1)}\,F_\Gcal(x/t) +{\rm const.}\,.
$$
Notice that the partition function does not change when we replace $(a,b,c,\kappa)$ with $(\lambda a, \lambda b, \lambda c, \lambda \kappa)$ (it is a homogeneous function of its arguments). Fixing $\kappa=1$ we can consider it as a symmetric function of $(a,b,c)$. 

In the following, we will define a ``refined" or ``extended" partition function as a function of $(a,b,c,t,\kappa)$, given by the same formulas above but with $t$ free, \ie, not fixed by \req{VogelRelation}. Note that we will still denote the partition function as $Z^{U}_\Gcal$. For particular choices of $(a,b,c,t,\kappa)$ we claim to reproduce refined Chern-Simons partition functions. 

\paragraph{$A_N$}

According to table \ref{VogelTable} one sets in the $A_{N-1}$ case 
$$
a=-2\,,\,\,\,\,\,b=2\,,\,\,\,\, c=t=N\,,
$$  
to recover the usual partition function of $SU(N)$ Chern-Simons on $\S^3$. Let us however set instead
\beq
\eqlabel{AnParas}
\boxed{
a=-2\,,\,\,\,\,\,b=2\beta \,,\,\,\,\, c=t=\beta N \,,\,\,\,\, \kappa=\kappa
}\,.
\eq  
The corresponding function $F_{A}(x)$ reads
\beq
\eqlabel{FA}
F_A(x)=\frac{\sinh\left(\frac{x(\beta N+1)}{2}\right)\sinh\left(\frac{x(\beta N-\beta)}{2}\right)}{\sinh\left(\frac{x}{2}\right)\sinh\left(\frac{x\beta}{2}\right)}-(N-1)(\beta N+1)\,.
\eq
The choice of parameters \req{AnParas} violates Vogel's condition \req{VogelRelation}. We claim however that this choice of parameters corresponds to refined $SU(N)$ Chern-Simons on $\S^3$. We will verify that this is indeed the case in section \ref{refCSAn}. The heuristic motivation for the parameters \req{AnParas} comes from the fact that this is the simplest choice yielding the expected denominator in a refined Chern-Simons integral representation out of \req{FGfull}. 

\paragraph{$D_N$}
Similarly, for refined $SO(2N)$ Chern-Simons on $\S^3$ we set
\beq
\eqlabel{DnParas}
\boxed{
a=-2\,,\,\,\,\,\,b=4\beta \,,\,\,\,\, c=2\beta N-4\beta\,,\,\,\,\,t=2\beta N-2\beta\,,\,\,\,\, \kappa=\kappa
} \,.
\eq  
The corresponding function $F_{D}(x)$ reads
\beq
\eqlabel{FD}
F_D(x)=\frac{2\sinh\left(\frac{x(2\beta N+1-2\beta)}{2}\right)\cosh\left(\frac{x(\beta N-2\beta)}{2}\right)\sinh\left(\frac{x\beta N}{2}\right)}{\sinh\left(\frac{x}{2}\right)\sinh\left(x\beta\right)}-N(2\beta N+1-2\beta)\,.
\eq
This can be as well written as
\beq
\eqlabel{FD2}
\begin{split}
F_D(x)=&\,\frac{\sinh\left(\frac{x(2\beta N+1-2\beta)}{2}\right)\sinh\left(\frac{x\beta(2N-2)}{2}\right)}{\sinh\left(\frac{x}{2}\right)\sinh\left(x\beta\right)}+\frac{\sinh\left(\frac{x(2\beta N+1-2\beta)}{2}\right)}{\sinh\left(\frac{x}{2}\right)}\\
&-N(2\beta N+1-2\beta)\,.
\end{split}
\eq
As for $A_N$, we claim that the choice of parameters \req{DnParas} corresponds to refined $SO(2N)$ Chern-Simons on $\S^3$. That this is indeed the case will be verified in section \ref{refCSDn}.

\section{Refined Chern-Simons}
\label{RCS}
\subsection{Generalities}
Following \cite{AS11,AS12}, the refined Chern-Simons partition function is defined as
\beq
\eqlabel{ZrCS}
Z^R_{\Gcal}:=c_\Gcal \prod_{m=0}^{\beta-1}\prod_{\alpha^+}\left(q^{-(\beta(\alpha,\rho)-m)/2}-q^{(\beta(\alpha,\rho)-m)/2} \right)\,,
\eq
where $c_\Gcal$ is a normalization factor, $\alpha^+$ denotes the set of positive roots and $\rho$ the Weyl vector of the root system of type $\Gcal$. Furthermore, we defined
\beq
q:=\exp{\frac{2\pi \ii}{\delta}}\,,
\eq
with $\delta:=\kappa+\beta h$ (this definition of $\delta$ coincides with \req{DeltaDef} for $t=\beta h$).  

It is more natural to write \req{ZrCS} as
$$
Z^R_\Gcal= (-2\ii)^{\beta|\alpha_+|} c_\Gcal \prod_{m=0}^{\beta-1}\prod_{\alpha^+}\sin\left(\pi \frac{\beta(\alpha,\rho)-m}{\delta} \right)\,.
$$
In analogy to the universal Chern-Simons partition function, we split the partition function into two parts, \ie,
$$
Z^R_\Gcal=Z^R_{\rm I,\Gcal}\, Z^R_{\rm II,\Gcal}\,,
$$ 
with
\beq
\eqlabel{ZrCSpertDef}
Z^R_{\rm I,\Gcal}:=\prod_{m=0}^{\beta-1}\prod_{\alpha^+}\frac{\left(\sin\left(\pi \frac{\beta(\alpha,\rho)-m}{\delta}\right) \right)}{\left(\pi\frac{(\beta(\alpha,\rho)-m)}{\delta}\right)}\,,
\eq
and
\beq
\eqlabel{ZrCSNPDef}
Z^R_{\rm II,\Gcal}:=c_\Gcal \left(-\frac{2\pi \ii}{\delta}\right)^{\beta|\alpha_+|} \prod_{m=0}^{\beta-1}\prod_{\alpha^+}\left(\beta(\alpha,\rho)-m\right)\,.
\eq

\paragraph{Part I}
The so-defined part I of the free energy $\Fcal^R_{\rm I,\Gcal}:=\log Z^R_{\rm I,\Gcal}$ can be easily evaluated. In detail, invoking the identity $\sin\pi x=\frac{\pi x}{\Gamma(1+x)\Gamma(1-x)}$ one infers
\beq
\begin{split}
\Fcal^R_{\rm I,\Gcal}&=\sum_{m=0}^{\beta-1}\sum_{\alpha^+}\log\frac{\sin\left(\pi \frac{\beta(\alpha,\rho)-m}{\delta}\right)}{\pi\frac{(\beta(\alpha,\rho)-m)}{\delta}}\\
&=-\sum_{m=0}^{\beta-1}\sum_{\alpha^+}\left(\log\Gamma\left(1+ \frac{\beta(\alpha,\rho)-m}{\delta} \right)+\log\Gamma\left(1-\frac{\beta(\alpha,\rho)-m}{\delta} \right)\right)\,.
\end{split}
\eq
With the integral representation (for $\Re z >0$) 
\beq
\eqlabel{logG}
\log\Gamma(z)=\int_{0}^\infty\frac{dx}{x}\frac{1}{e^{x}-1}\left( (z-1)(1-e^{-x})+e^{-x(z-1)}-1\right)\,,
\eq
this becomes
\beq
\eqlabel{FrCSpertInt}
\Fcal^R_{\rm I,\Gcal}=-\int_0^\infty\frac{dx}{x}\frac{1}{e^{x}-1}\sum_{m=0}^{\beta-1}\sum_{\alpha^+}\left(e^{-x\frac{m}{\delta}}e^{x \frac{\beta(\alpha,\rho)}{\delta}}-e^{x\frac{m}{\delta}}e^{-x \frac{\beta(\alpha,\rho)}{\delta}}-2\right)\,.
\eq
Let us define for notational simplicity $\Xi(x,\beta):=\sum_{m=0}^{\beta-1}e^{-x\frac{m}{\delta}}$ and $p:=e^{\frac{x \beta}{\delta}}$. With these definitions, the summation part of the integrand in \req{FrCSpertInt} reads
\beq
\eqlabel{FrCSpertIntroots}
\boxed{
\sum_{\alpha^+}\left(\Xi(x,\beta)\,p^{(\alpha,\rho)}-\beta \right)+\sum_{\alpha^-}\left(\Xi(-x,\beta)\,p^{-(\alpha,\rho)}-\beta \right)
}\,.
\eq
Since we have $\Xi(x,1)=1$, the combined summations over the roots can be performed in the $\beta=1$ case using the results of \cite{MV12}. However, we do not know at present how to perform the individual summations in general, hence we have to evaluate the integrand \req{FrCSpertIntroots} in a case by case basis, as we will do for $A_N$ in section \ref{refCSAn} and for $D_N$ in section \ref{refCSDn}. For that, it will be useful to note that we have in terms of $p$
\beq
\eqlabel{Xiviap}
\begin{split}
\Xi(\pm x,\beta)&=p^{\pm(1/\beta-1)/2} \frac{p^{1/2}-p^{-1/2}}{p^{1/(2\beta)}-p^{-1/(2\beta)}} \,.
\end{split}
\eq

\paragraph{Part II}
Let us now consider part II given in \req{ZrCSNPDef}. This part can be written as
$$
\Fcal^R_{\rm II,\Gcal}=\log c_S+\beta|\alpha_+|\log\left(-\frac{2\pi\ii}{\delta}\right)+ \sum_{\alpha^+}\left(\log\left(1+\beta(\alpha,\rho)\right)-\log\left(1-\beta+\beta(\alpha,\rho)\right)  \right)\,.
$$
Similarly as for part I, we can invoke the integral representation of $\log\Gamma$ given in \req{logG} to rewrite the non-constant part of $\Fcal_{\rm II}$ as the integral $\int_{0}^\infty\frac{dx}{x}\frac{1}{e^{x}-1}$ over
\beq
\eqlabel{Fnpintgeneral}
\begin{split}
\boxed{
\beta |\alpha^+|(p^{\delta/\beta}-1)p^{-\delta/\beta}-\left(p^\delta-1\right)\sum_{\alpha^+}p^{-\delta(\alpha,\rho)}
}\,.
\end{split}
\eq
As for part I we do not know how to perform the left-over summation over the positive roots in general and have to proceed case by case, as we will do in the following sections.

\subsection{$A_N$}
\label{refCSAn}
For $SU(N)$ (corresponding to $A_{N-1}$) we have $\frac{1}{2}N(N-1)$ positive roots. A basis is given by $e_I-e_J$ with $I<J$ and $(e_I)_j=\delta_{I,j}$ such that for a vector $v$ we have that $(e_I-e_J,v)=v_i-v_j$. The Weyl vector $\rho$ is given by $(\rho)_j=(N+1)/2-j$. Hence, $(\alpha,\rho)=e_J-e_I$ and $\prod_{\alpha^+}=\prod_{I<J}^N$. 

Thus we have,
\beq
\eqlabel{Anpsum}
\begin{split}
\sum_{\alpha^+}p^{(\alpha,\rho)}&=-\frac{Np}{(p-1)}+\frac{p(p^{N}-1)}{(p-1)^2}\,,\\
\sum_{\alpha^+}p^{-(\alpha,\rho)}&=\frac{N}{(p-1)}-\frac{p(1-p^{-N})}{(p-1)^2}\,.
\end{split}
\eq

\paragraph{Part I}
This leads to the following explicit expression of \req{FrCSpertIntroots} for $A_{N-1}$ root system
\beq
\eqlabel{AnPartIint}
\begin{split}
\frac{(\Xi(x,\beta)p^{N/2} -\Xi(-x,\beta)p^{-N/2})(p^{N/2}-p^{-N/2}) }{(p^{1/2}-p^{-1/2})^2}+N\frac{\Xi(-x,\beta)-p\Xi(x,\beta)}{p-1}-\beta N(N-1)
\end{split}
\eq
Invoking \req{Xiviap}, one infers that the integrand \req{AnPartIint} can be written as 
\beq
\eqlabel{FAdeltaShift}
\frac{\sinh\left(\frac{x(\beta N-\beta+1)}{2\delta}\right)\sinh\left(\frac{xN\beta}{2\delta}\right) }{\sinh\left(\frac{x\beta}{2\delta}\right)\sinh\left(\frac{x}{2\delta}\right)}-N(\beta N+1-\beta)\,.
\eq
With the trigonometric identity \req{TrigId1}, we deduce that this equals precisely $F_A(x/\delta)$ given in \req{FA}. Thus, we have shown that
$$
\boxed{
Z^{R}_{{\rm I},A}= \exp \int_0^\infty\frac{dx}{x(e^{x}-1)}\, F_A(x/\delta)=Z^{U}_{{\rm I},A}
}\,.
$$

\paragraph{Part II}
Invoking \req{Anpsum}, the integrand of part II given in \req{Fnpintgeneral} reads for $A_{N-1}$ root system
\beq
\eqlabel{AnNPint2}
\begin{split}
&\frac{1}{2}\beta N(N-1)(p^{\delta/\beta}-1)p^{-\delta/\beta}+\left(p^{\delta}-1\right)\left(\frac{p^{\delta}(1-p^{-\delta N})}{(p^\delta-1)^2}-\frac{N}{p^\delta-1}\right)\,.\\
\end{split}
\eq
It is convenient to factor out $(p^{\delta/\beta}-1)=(e^{x}-1)$ such that the integration reduces to $\int_0^\infty\frac{dx}{x}$ with integrand
\beq
\eqlabel{AnPIIint1}
\frac{1}{2}\beta N(N-1)p^{-\delta/\beta}+\frac{p^{\delta}(1-p^{-\delta N})}{(p^{\delta/\beta}-1)(p^\delta-1)}-\frac{N}{p^{\delta/\beta}-1}\,.
\eq
We want to show that this integral equals $\int_0^\infty \frac{dx}{x(e^x-1)} F_A(x/t)+{\rm const.}$ . Therefore, using that for $A_{N-1}$ we have $t=\beta N$, we factor out $(e^x-1)$ from $F_A$ such that the integration changes to $\int_0^\infty\frac{dx}{x}$. This then allows us to further rescale $x\rightarrow  t x$. Hence, the integrand can be written as 
\beq
\eqlabel{AnPIIint2}
\frac{p^{\delta/\beta}(1-p^{-\delta(N+1/\beta-1)})}{(p^{\delta}-1)(p^{\delta/\beta}-1)}-\frac{N(\beta N+1-\beta)}{p^{\delta N}-1}\,.
\eq
Substracting \req{AnPIIint2} from \req{AnPIIint1} yields
$$
\frac{1}{2}\beta N(N-1)p^{-\delta/\beta}-\frac{N}{p^{\delta/\beta}-1}+\frac{N(\beta N+1-\beta)}{p^{\delta N}-1}+\frac{p^{\delta}-p^{\delta/\beta}}{(p^\delta-1)(p^{\delta/\beta}-1)}\,,
$$
which need to integrate to a constant under $\int_{0}^\infty\frac{dx}{x}$. In terms of the integrals of appendix \ref{IntApp} this integral reads
$$
\lim_{\ep\rightarrow 0}\left( \frac{1}{2}\beta N(N-1)\,\Ical_1(1)+N(\beta N+1-\beta)\,\Ical_2(\beta N,\beta N) -N\, \Ical_2(1,1)+ 
\Ical_3(\beta,1)\right)\,,
$$
and can be evaluated to
$$
\frac{\beta N(N-1)}{2}\log \frac{\beta N}{2\pi}+\frac{N}{2} \log \beta N-\log \sqrt{\beta} \,.
$$
Hence, we have that 
$$
\boxed{
Z^R_{{\rm II},A}= \frac{c_A\, t^{\frac{N}{2}}}{\sqrt{\beta}}\left(-\ii\frac{t}{\delta}\right)^{\frac{\beta N(N-1)}{2}} \exp \int_0^\infty \frac{dx}{x(e^x-1)}\,F_A(x/t)\sim Z^U_{{\rm II},A}}\,.
$$
This completes the proof of the claim for $A_N$ root system.

\subsection{$D_N$}
\label{refCSDn}
For $SO(2N)$ (corresponding to $D_N$ root system) we have $N(N-1)$ positive roots. A basis is given by $e_I-e_J, e_I+e_J$ with $I<J$. The Weyl vector reads $(\rho)_i=N-i$\,. We have $(e_I-e_J,\rho)=J-I$ and $(e_I+e_J,\rho)=2N-I-J$.

Hence, 
\beq
\eqlabel{Dnpsum}
\begin{split}
\sum_{\alpha^+}p^{(\alpha,\rho)}&=\frac{p(p^N-1)(p^{N-1}+p)}{(p-1)^2(p+1)}-\frac{N p}{p-1}\,,\\
\sum_{\alpha^+}p^{-(\alpha,\rho)}&=-\frac{p(p^N-1)(p^{-(2N-2)}+p^{-N})}{(p-1)^2(p+1)}+\frac{N}{p-1}\,.
\end{split}
\eq

\paragraph{Part I}
We infer that \req{FrCSpertIntroots} reads for $D_N$
$$
\frac{p^{3/2-2 N-(1/\beta-1)/2}(p^N-1)(p^{N-2}+1)(p^{2N+1/\beta-2}-1)}{(p^{1/(2\beta)}-p^{-1/(2\beta)})(p^{1/2}-p^{-1/2})(p^{1/2}+p^{-1/2})}-N(2\beta N+1-2\beta)\,.
$$ 
Or, in terms of trigonometric functions
$$
\frac{\sinh\left(\frac{x \beta N}{2\delta}\right) \cosh\left(\frac{x(\beta N-2\beta) }{2\delta}\right)\sinh\left(\frac{x (2\beta N+1-2\beta)}{2\delta}\right)}{\sinh\left(\frac{x}{2\delta}\right) \sinh\left(\frac{ x\beta}{2\delta}\right)\cosh\left(\frac{ x\beta}{2\delta}\right)}-N(2\beta N+1-2\beta)\,,
$$
which can be matched with $F_D(x/\delta)$ given in \req{FD}. Thus,
$$
\boxed{
Z^R_{{\rm I},D}=\int_0^\infty \frac{dx}{x(e^{x}-1)}\, F_D(x/\delta)= Z^U_{{\rm I},D}
}\,.
$$

\paragraph{Part II}
Similarly, using \req{Dnpsum} we obtain in the $D_N$ case for \req{Fnpintgeneral} the expression
\beq
\eqlabel{DnNPint2}
\begin{split}
\beta N(N-1)(p^{\delta/\beta}-1)p^{-\delta/\beta}+\frac{p^\delta(p^{\delta N}-1)(p^{-\delta(2N-2)}+p^{-\delta N})}{(p^\delta-1)(p^\delta+1)}-N\,.\\
\end{split}
\eq
As in the $A_N$ case, it is convenient to factor out $(e^x-1)$ such that we have the integral $\int_0^\infty\frac{dx}{x}$ with integrand
$$
\beta N(N-1)p^{-\delta/\beta}+\frac{p^{\delta}(p^{\delta N}-1)(p^{-\delta(2N-2)}+p^{-\delta N})}{(p^{\delta/\beta}-1)(p^{2\delta}-1)}-\frac{N}{(p^{\delta/\beta}-1)}\,.
$$
It is useful to split the second term further, yielding
\beq
\eqlabel{DnNPint3}
\beta N(N-1)p^{-\delta/\beta}+\frac{p^{\delta }(1-p^{-\delta (2N-2)})}{(p^{\delta/\beta}-1)(p^{2\delta}-1)}+\frac{p^{-\delta (N-1)}}{(p^{\delta/\beta}-1)}-\frac{N}{(p^{\delta/\beta}-1)}\,.
\eq

Similarly, the integral $\int_0^\infty\frac{dx}{x(e^x-1)}F_D(x/t)$ (\cf, \req{FD2}), can be rewritten as the integral $\int_0^\infty\frac{dx}{x}$ over (with a redefinition $x\rightarrow t x$)
\beq
\eqlabel{DnNPint4}
\frac{p^{\delta(1+1/\beta)}(1-p^{-\delta(2N-2+1/\beta)})}{(p^{\delta/\beta}-1)(p^{2\delta}-1)}+\frac{p^{-\delta t/(2\beta)}(p^{\delta(t+1)/\beta}-1)}{(p^{t\delta/\beta}-1)(p^{\delta/\beta}-1)}-\frac{N(2\beta N+1-2\beta)}{p^{t\delta/\beta}-1}\,.
\eq
Substracting \req{DnNPint4} from \req{DnNPint3} yields the left-over terms
\beq
\begin{split}
&\frac{Nt}{2}p^{-\delta/\beta}-\frac{N}{(p^{\delta/\beta}-1)}+\frac{p^{-\delta (N-1)}}{(p^{\delta/\beta}-1)}+\frac{N(t+1)}{p^{t\delta/\beta}-1}- \frac{p^{-\delta t/(2\beta)}(p^{\delta(t+1)/\beta}-1)}{(p^{t\delta/\beta}-1)(p^{\delta/\beta}-1)}-\frac{p^{\delta}}{(p^{2\delta}-1)}\,.
\end{split}
\eq
In terms of the integrals of appendix \ref{IntApp} this reads
\beq
\begin{split}
\lim_{\ep\rightarrow 0}&\left(\frac{Nt}{2}\,\Ical_1(1)-N\,\Ical_2(1,1)+\Ical_2(1,1+t/2)+N(t+1)\,\Ical_2(t,t)-\Ical_4(t,1,t/2)-\Ical_5(2\beta)\right)\,,
\end{split}
\eq
and evaluates to
$$
N(\beta N-\beta)\log \left(\frac{2\beta N-2\beta}{2\pi }\right)+\frac{N}{2}\log\left(2\beta N-2\beta\right)-\log 2\,.
$$
Hence, 
$$
\boxed{
Z^R_{{\rm II},D}=\frac{c_D\, t^{\frac{N}{2}}}{2}\left(-\ii\frac{t}{\delta}\right)^{\beta N(N-1)}\exp \int_0^\infty \frac{dx}{x(e^{x}-1)}\, F_D(x/t)\sim Z^U_{{\rm II},D}
}\,.
$$
This completes the proof in the $D_N$ case.

\section{Large $N$ expansion}
\label{largeN}
It is instructive to perform the large $N$ expansion of the integral representation in the $SU(N)$ case. For that, we keep fixed at large $N$
\beq
\eqlabel{ParasLargeN}
g_s:=\frac{1}{\delta}\,,\,\,\,\,\,\mu:=\frac{t}{\delta}= g_s t\,.
\eq

The function $F_A(x/\delta)$ defined in \req{FA} reads in terms of the parameters \req{ParasLargeN}
$$
F_A(x,\mu,g_s;\beta )=\frac{\sinh\left(\frac{x(\mu+g_s)}{2}\right)\sinh\left(\frac{x (\mu-\beta g_s)}{2}\right)}{\sinh\left(\frac{x g_s}{2}\right)\sinh\left(\frac{x\beta g_s}{2}\right)}-\frac{(\mu-\beta g_s)(\mu+g_s )}{\beta g^2_s}\,.
$$
It is more convenient and natural to perform the additional quantum shift
$$
\mu\rightarrow \mu'=\mu-\frac{1}{2}(1-\beta)g_s\,,
$$
in the definition of $\mu$ such that
$$
F_A(x,\mu',g_s;\beta )=\frac{\sinh\left(\frac{x(2\mu+(1+\beta)g_s)}{4}\right)\sinh\left(\frac{x (2\mu-(1+\beta) g_s)}{4}\right)}{\sinh\left(\frac{x g_s}{2}\right)\sinh\left(\frac{x\beta g_s}{2}\right)}-\frac{(\mu^2-(1+\beta)^2 g^2_s)}{4\beta g^2_s}\,.
$$

\paragraph{Part I}

Under this redefinition, $F_A$ possesses an expansion into even powers of $g_s$ and $\mu$ only, \ie,
\beq
\eqlabel{FAexpansion}
\begin{split}
F_A(x,\mu',g_s;\beta)&=\frac{2}{\beta g_s^2}\sum_{k=1}^\infty \frac{x^{2k} \mu^{2k+2}}{(2k+2)!}+ 2\,\Psi_A^{(0)}(\beta)\sum_{k=1}^\infty \frac{(x\mu)^{2k}}{(2k)!}\\
&+2\sum_{g=2}^\infty (\sqrt{\beta} g_s)^{2g-2}\,\left(\Psi_A^{(2g-2)}(\beta)-\Phi_A^{(2g-2)}(\beta)\right) \frac{x^{2g-2}}{(2g-3)!}\\
&+2\sum_{g=2}^\infty (\sqrt{\beta} g_s)^{2g-2}\,\frac{\Psi_A^{(2g-2)}(\beta) }{(2g-3)!}\sum_{k=1}^\infty\frac{x^{2(g+k-1)} \mu^{2k}}{(2k)!}\,,
\end{split}
\eq
with coefficients
\beq
\eqlabel{DefPhiA}
\begin{split}
\Phi_A^{(0)}(\beta)&:=\frac{1}{12}\left(\beta+\frac{1}{\beta}\right)-\frac{1}{4}\,,\\
\Phi_A^{(n>0)}(\beta)&:=(n-1)!\sum_{k=0}^{n+2}(-1)^k\frac{B_{k}B_{n+2-k}}{k!(n+2-k)!}\,\beta^{k-n/2-1}\,,\\
\end{split}
\eq
and
\beq
\begin{split}
\Psi_A^{(0)}(\beta)&:=-\frac{1}{24}\left(\beta+\frac{1}{\beta}\right)\,,\\
\Psi_A^{(n>0)}(\beta)&:=(n-1)!\sum_{k=0}^{n+2}(-1)^k\frac{B_{k}B_{n+2-k}}{k!(n+2-k)!}(2^{1-k}-1)(2^{k-n-1}-1)\,\beta^{k-n/2-1}\,.
\end{split}
\eq
The latter coefficients correspond to the expansion coefficients of the free energy of the $c=1$ string at radius $R\propto \beta$, originally derived in \cite{GK90}.

Invoking the Mellin-transform of the Riemannian $\zeta$-function 
\beq
\eqlabel{MellinRZ}
\Gamma(s)\zeta(s)=\int_0^\infty \frac{dx}{x}\frac{x^{s}}{e^{x}-1}\,,
\eq
we deduce from \req{FAexpansion} that at large $N$ the part I of the free energy reads
\beq
\eqlabel{FIAgsDef}
\Fcal_{{\rm I},A}(\mu',g_s;\beta)\sim \sum_{g=0}^\infty \Fcal^{(g)}_{{\rm I},A}(\mu;\beta)\,(\sqrt{\beta}g_s)^{2g-2}\,,
\eq
with
\beq
\begin{split}
\Fcal^{(0)}_{{\rm I},A}(\mu;\beta)=&2\sum_{k=1}^\infty \frac{\zeta(2k)}{2k(2k+1)(2k+2)} \,\mu^{2k+2}\,,\\
\Fcal^{(1)}_{{\rm I},A}(\mu;\beta)=&\,\Psi^{(0)}(\beta) \sum_{k=1}^\infty\frac{\zeta(2k)}{k} \mu^{2k}\,,\\
\Fcal^{(g>1)}_{{\rm I},A}(\mu;\beta)=&\,2\left(\Psi_A^{(2g-2)}(\beta)-\Phi_A^{(2g-2)}(\beta)\right)\,\zeta(2g-2)\\
&+2\frac{\Psi_A^{(2g-2)}(\beta) }{(2g-3)!}\sum_{k=1}^\infty\frac{\zeta(2(g+k-1))\Gamma(2(g+k-1)) }{(2k)!}\,\mu^{2k}\,.
\end{split}
\eq
Since for $\beta=1$ we have
$$
\Phi_A^{(n)}(1)=\Psi_A^{(n)}(1)=\chi^{(n)}\,,
$$
where $\chi^{(n)}$ denotes the virtual Euler characteristic of the moduli space of genus $n$ complex curves, we precisely recover at large $N$ the ``perturbative" free energy of the topological string on the resolved conifold, see \cite{GV98,GV981}. Moreover, we observe that the $\beta$ degree of freedom of the universal Chern-Simons partition function introduced in \req{AnParas}, or equivalently refinement, essentially corresponds to the replacement
\beq
\eqlabel{NewEuler}
\boxed{\chi^{(n)}\rightarrow \Psi^{(n)}(\beta)}\,,
\eq
at large $N$. This explicitly confirms earlier results of \cite{KW10a}. In particular, one may see the replacement \req{NewEuler}, that is, refinement, as substituting $\chi$ with the parameterized Euler characteristic of \cite{GHJ01} (\cf, \cite{KW12}).

\paragraph{Part II}
The essential difference between part I and II of the partition function is a mere substitution $\delta\rightarrow t$. Therefore we set $\mu=1$ in \req{FAexpansion}, as is clear from \req{ParasLargeN}. Hence, we have to integrate
\beq
\begin{split}
F_A(x,\mu'-\mu+1,g_s;\beta)&=\frac{1}{\beta g_s^2}\left(\frac{2\cosh(x)-x^2-2}{x^2}\right)+2\,\Psi_A^{(0)}(\beta)(\cosh(x)-1)\\
&+2\sum_{g=2}^\infty (\sqrt{\beta} g_s)^{2g-2}\,\left(\Psi_A^{(2g-2)}(\beta)-\Phi_A^{(2g-2)}(\beta)\right) \frac{x^{2g-2}}{(2g-3)!}\\
&+2\sum_{g=2}^\infty (\sqrt{\beta} g_s)^{2g-2}\,\frac{\Psi_A^{(2g-2)}(\beta) }{(2g-3)!}\,x^{2g-2}(\cosh(x)-1)\,.
\end{split}
\eq
Defining $\Fcal_{{\rm II},A}$ as in \req{FIAgsDef}, and using the Mellin-transform of the Hurwitz $\zeta$-function 
\beq
\eqlabel{MellinHZ}
\Gamma(z)\zeta(z,v/w)=w^z\int_{0}^\infty \frac{dx}{x}\, \frac{x^ze^{(w-v)x}}{e^{wx}-1}\,,
\eq
we infer
\beq
\begin{split}
\Fcal_{{\rm II},A}^{(0)}(\beta)&=\frac{1}{2}\log 2\pi -\frac{3}{4}\,,\\
\Fcal_{{\rm II},A}^{(1)}(\beta)&=-\Psi_A^{(0)}(\beta)\,\lim_{\ep\rightarrow 0}\,\Ical_1(\ep)\,,\\
\Fcal_{{\rm II},A}^{(g>1)}(\beta)&=2\left(\Psi_A^{(2g-2)}(\beta)-\Phi_A^{(2g-2)}(\beta)\right)\zeta(2g-2)-\Psi_A^{(2g-2)}(\beta)\,,
\end{split}
\eq
with the 1-loop (order $g_s^0$) logarithmic divergence given by \req{I1}. At $\beta=1$ this reproduces the results of \cite{OV02} obtained via an asymptotic expansion of Barnes $G$-function (up to the subtile logarithmic divergence at 1-loop).

\acknowledgments{

The work of D.K. has been supported by a Simons fellowship, and by the Berkeley Center for Theoretical Physics. The work of A.S. was supported by a NSF grant.
}
\appendix

\section{Integrals}
\label{IntApp}
In this appendix some not so common integrals needed for the calculations in the main text are derived.

\subsection{$\Ical_1(a>0):=\int_\ep^\infty \frac{dx}{x} e^{- a x}$}
We have
\beq
\eqlabel{I1}
\boxed{
\Ical_1(a)=-\log \ep -\log a -\gamma+\Ocal(\ep)
}\,,
\eq
with $\gamma$ the Euler-Mascheroni constant.

\subsection{$\Ical_2(a>0,b\geq a)=\int_\ep^\infty\frac{dx}{x}\frac{e^{(a-b)x}}{e^{ax}-1}$}
\label{AppAInt2}
First, note that
$$
\frac{1}{e^{a x}-1}=\frac{1}{a x}+\frac{1-e^{a x}+ax}{ax(e^{ax}-1)}\,.
$$
Hence,
\beq
\eqlabel{FundIntegrand1}
\frac{e^{(a-b)x}}{(e^{a x }-1)}=\frac{e^{(a-b)x}}{a x}-\frac{a x e^{(a-b)x}}{2(e^{a x}-1)}-\frac{e^{(a-b)x}}{(e^{a x}-1)}\sum_{n=3}^\infty\frac{(ax)^{n-1}}{n!}\,.
\eq
With the help of the Mellin-transform \req{MellinHZ} we then infer from \req{FundIntegrand1} that
\beq
\begin{split}
\lim_{\ep\rightarrow 0}\Ical_2(a>0,b\geq a)=&\,\frac{1}{a \ep}+\left(\frac{b}{a}-1\right)(\log\ep+\gamma-1+\log(b-a))\\
&+\frac{1}{2}\left(\log\ep +\log a+\gamma+\psi(b/a)\right)-\sum_{n=2}^\infty \frac{\zeta(n,b/a)}{n(n+1)}\,,
\end{split}
\eq
with $\psi(x)$ the Digamma function and $\zeta(w,v)$ the Hurwitz $\zeta$-function.

In particular, 
$$
\boxed{
\lim_{\ep\rightarrow 0}\Ical_2(a>0,b=a)=\frac{1}{a\ep}+\frac{1}{2}\log \ep+\frac{1}{2}\left(\log a+\gamma-\log 2\pi\right)}\,,
$$
and for $a,b$ integer
\beq
\eqlabel{I2aba}
\boxed{
\begin{split}
\lim_{\ep\rightarrow 0}\Ical_2(a,a b\geq a)=&\,\frac{1}{a\ep}+(b-1/2)\log\ep+(b-1/2)\log a+(b-1/2)\gamma\\
&+\log\Gamma(b) -\frac{1}{2}\log 2\pi
\end{split}
}\,,
\eq
where we used for $\Ical_2(a,a b)$ the relation
$$
\sum_{n=2}^\infty\frac{\zeta(n,1+c)}{n(n+1)}=-\frac{1}{2}\gamma+\frac{1}{2}\log 2\pi+\frac{1}{2}\left(H_c-2c\right)+\log\frac{c^{c-1}}{(c-1)!}\,,
$$
with $H_n$ the $n$th Harmonic number and $c$ an integer.

Further, note that $\sum_{n=2}^\infty\frac{\zeta(n,3/2)}{n(n+1)}=\frac{1-\gamma}{2}$ and $\psi(3/2)=-\gamma-2\log 2+2$. Hence,
\beq
\eqlabel{I2w32}
\boxed{
\lim_{\ep\rightarrow 0}\Ical_2(a,3/2a)=\frac{1}{a\ep}+\log\ep+\log a+\gamma-\frac{3}{2}\log 2
}\,.
\eq

\subsection{$\Ical_3(a>0,b>0):=\int_\ep^\infty \frac{dx}{x}\frac{e^{ax}-e^{bx}}{(e^{a x}-1)(e^{b x}-1)}$}
Note that
$$
\frac{e^{a x}-e^{b x}}{(e^{a x}-1)(e^{b x}-1)}=\frac{1}{e^{bx}-1}-\frac{1}{e^{ax}-1}\,.
$$
Hence,
$$
\Ical_3(a>0,b>0)=\Ical_2(b,b) -\Ical_2(a,a) \,.
$$
In particular,
\beq
\begin{split}
\boxed{
\lim_{\ep\rightarrow 0}\Ical_3(a,1)=\left(1-\frac{1}{a}\right)\frac{1}{\ep}-\frac{1}{2}\log a
}\,.
\end{split}
\eq
\subsection{$\Ical_4(a>0,b>0,c>0):=\int_\ep^\infty \frac{dx}{x}\frac{(e^{(a+b)x}-1)e^{-cx}}{(e^{a x}-1)(e^{b x}-1)}$}
We have
$$
\frac{e^{(a+b)x}-1}{(e^{a x}-1)(e^{b x}-1)}=\frac{1}{e^{ax}-1}+\frac{e^{bx}}{e^{bx}-1}\,.
$$
Hence,
$$
\Ical_4(a,b,c)=\Ical_2(a,a+c)+\Ical_2(b,c)\,.
$$
Invoking \req{I2aba} and \req{I2w32} we deduce for $a$ even
$$
\boxed{
\lim_{\ep\rightarrow 0}\Ical_4(a,1,a/2)=\left(1+\frac{1}{a}\right)\frac{1}{\ep}+\frac{a+1}{2}\left(\log\ep+\gamma\right)+\log a+\log\Gamma(a/2)-2\log 2-\frac{1}{2}\log\pi
}\,.
$$

\subsection{$\Ical_5(a):=\int_\ep^\infty \frac{dx}{x} \frac{e^{a x/2}}{e^{a x}-1}$}
We write
$$
 \frac{e^{a x/2}}{e^{a x}-1}=\frac{1}{e^{a x}-1}+\frac{a x}{2(e^{a x}-1)}+\frac{1}{e^{a x}-1}\sum_{n=2}^\infty \frac{(a x)^n}{2^n\, n!}\,.
$$
Hence,
$$
\lim_{\ep\rightarrow 0}\Ical_5(a)=\Ical_2(a,a)-\frac{1}{2}\log\ep-\frac{1}{2}\log a+\sum_{n=2}^\infty \frac{\zeta(n)}{2^n\, n}\,.
$$
Using that $\sum_{n=2}^\infty \frac{\zeta(n)}{2^n\, n}=-\frac{1}{2}\gamma+\frac{1}{2}\log\pi$, we deduce
$$
\boxed{
\lim_{\ep\rightarrow \ep}\Ical_5(a)=\frac{1}{a\ep}-\frac{1}{2}\log 2
}\,.
$$

\section{Trigonometric identity}
\label{TrigIdapp}
In this section we will derive the trigonometric identity 
\beq
\eqlabel{TrigId1}
\boxed{
\sinh(c (x-y+1))\sinh(c x) -\sinh(c y)\sinh(c) = \sinh(c(x-y))\sinh(c(x+1))
}\,.
\eq
The proof of this identity immediately follows via invoking the fundamental identity 
$$
\sinh(x)\sinh(y)=\frac{1}{2}\left(\cosh(x+y)-\cosh(x-y)\right)\,.
$$
Namely,
\beq
\begin{split}
&\sinh(c (x-y+1))\sinh(c x)-\sinh(c y)\sinh(c) \\
&=\frac{1}{2}\left(\cosh(c(2x-y+1)-\cosh(c(y-1))- \cosh(c(y+1)+\cosh(c(y-1))  \right)\\
&=\frac{1}{2}\left(\cosh(c(2x-y+1)- \cosh(c(y+1)\right)\\
&=\sinh(c(x-y))\sinh(c(x+1))\,.
\end{split}
\eq

\end{document}